\documentclass[11pt]{article}
\usepackage{epsfig,sint,cite,amsfonts}
\usepackage{epsfig}
\usepackage{latexsym}
\usepackage{amssymb}
\usepackage{amsmath}
\usepackage{booktabs}
\usepackage{graphicx}% Include figure files
\usepackage{dcolumn}% Align table columns on decimal point

\usepackage[english]{babel}
\usepackage{bm}% bold math

\renewcommand{\vec}[1]{\boldsymbol{#1}}
\newcommand{\be}{\begin{equation}}
\newcommand{\ee}{\end{equation}}
\newcommand{\ba}{\begin{eqnarray}}
\newcommand{\ea}{\end{eqnarray}}
\newcommand{\la}{\label}

\newcommand{\bn}{\boldsymbol{n}}

\newcommand{\bx}{\boldsymbol{x}}
\newcommand{\bk}{\boldsymbol{k}}

\newcommand{\ud}{\,\mathrm{d}}
\newcommand{\half}{{\textstyle\frac{1}{2}}}

\newcommand{\<}{\langle}
\renewcommand{\>}{\rangle}
\newcommand{\nn}{\nonumber \\}

\newcommand{\txts}{\textstyle}
\newcommand{\eq}{Eq.~}

\begin{document}
\begin{flushright}
MITP/13-017 
\end{flushright}

\begin{center}
{\LARGE\bf Lattice QCD and the two-photon decay \\ of the neutral pion \\[0.5ex]} 
\end{center}
\vskip 1.6cm

\centerline{\Large Harvey B.\ Meyer}
\vskip 0.9cm
\noindent\hspace{-0.15cm} \emph{PRISMA Cluster of Excellence, 
Institut f\"ur Kernphysik and Helmholtz Institute Mainz, Johannes Gutenberg-Universit\"at Mainz, D-55099 Mainz}

\vspace{1.6cm}
\centerline{\bf Abstract}
\vspace{0.3cm}

Two-photon decays probe the structure of mesons and represent an
important contribution to hadronic light-by-light scattering. For the
neutral pion, the decay amplitude tests the effects of the chiral
anomaly; for a heavy quarkonium state, it measures the magnitude of
its wavefunction at the origin.  We rederive the expression of the
decay amplitude in terms of a Euclidean correlation function starting
from the theory defined on the torus.  The derivation shows that for
timelike photons the approach to the infinite-volume decay amplitude
is exponential in the periodic box size.

\vspace{1.6cm}

\section{Introduction}

Photon-hadron interactions constitute a venerable subject that
continues to be important in 21st century particle physics.
Historically, the inner structure of the proton was investigated in
elastic scattering experiments of electromagnetic
probes~\cite{PhysRev.98.217}. Roughly a decade later, the study of
deeply inelastic
processes~\cite{PhysRevLett.23.930,PhysRevLett.23.935} led to the
development of the parton model, thereby making a decisive
contribution to the advent of QCD as the theory of the strong
interactions. At the same time, the decay $\pi_0\to\gamma\gamma$ led
to the realization that the flavor-singlet chiral symmetry of the
classical field theory is broken by quantum
effects~\cite{Adler:1969gk,Bell:1969ts}.  In this article we rederive
the expression for the hadronic matrix element that determines the
two-photon width of the neutral pion, starting from QCD defined on a
finite torus. This is the context in which almost all lattice QCD
calculations are performed. An important conclusion of this analysis
is that the finite-size effects on the type of hadronic matrix
elements that several lattice QCD collaborations are
computing~\cite{Feng:2012ck,Lin:2013im,Feng:2012an} are exponentially
suppressed.

We assume that QCD has exact isospin symmetry prior to coupling
hadrons to photons. We consider a neutral, pseudoscalar meson of mass
$M$ which is stable in QCD and which we refer to as the `pion',
although we need not assume that it is a pseudo-Goldstone boson.  The
basic idea is to think of the square module of the decay amplitude as
the pion pole contribution to the light-by-light scattering amplitude.
We will be led to consider a center-of-mass energy off by order $e^2$
from the pion pole ($e$ is the electromagnetic coupling constant). At
this energy, the $\gamma\gamma$ cross-section is of order $e^2$ rather
than $e^4$. For the sake of the argument, we will consider massive
vector bosons $v$ with a finite mass $M_v\leq M/2$ and denote the
corresponding field theory by mQED.  The QCD part of the $\pi_0$ decay
amplitude into two timelike photons is reinterpreted as the QCD part
of the $\pi_0$ decay amplitude into two massive, on-shell vector bosons.  We
can then use the general formalism developed by
L\"uscher~\cite{Luscher:1986pf,Luscher:1990ux} to establish a
correspondence between the (discrete) spectrum of mQED+QCD on the
torus and the $vv$ scattering phase. The $\pi_0-vv$ interaction
leads to a splitting of the $\pi_0$ mass and a two-vector-boson energy
level, given by matrix elements of the finite-volume Hamiltonian. The
relevant matrix element can then be rewritten in terms of a
time-ordered product of vector currents in the finite-volume theory.
Since the relation between the finite-volume spectrum and the
scattering amplitude holds up to exponentially suppressed corrections,
we find that this conclusion also holds for the module of the
$\pi_0\to\gamma^*\gamma^*$ amplitude.

Once the finite-volume corrections have been treated in this way for
timelike photons, it is a seemingly innocuous step to extend the
conclusion about the finite-size effects to the case of real
(lightlike) photons. Rather than giving a formal proof, we check that
the relation between the finite-volume matrix element and the
infinite-volume matrix element holds even for massless photons in two
simples field theories.

One might at first think that, photons being massless, the
finite-volume effects will only be suppressed by a power of the linear
extent of the torus. However, at leading non-trivial order in the
electromagnetic coupling, the photons are merely external particles in
the reaction of interest. Their interaction is of purely hadronic
origin, and has a range of order the inverse pion Compton
wavelength. Inelastic channels are kinematically allowed; but they are
suppressed by the smallness of the electromagnetic coupling.  We note
that the partial wave expansion which underlies L\"uscher's formalism
is also valid for relativistic and even massless particles (see for
instance~\cite{Weinberg:1995mt}, section 3.7).

Our treatment of the problem bears a strong similarity with the
formulation of the $K\to\pi\pi$ decay on the torus by Lellouch and
L\"uscher~\cite{Lellouch:2000pv}.  In some ways, the
$\pi_0\to\gamma\gamma$ amplitude is simpler: there are no final-state
interactions at order $e^2$. In other ways, it is more complicated:
the decay amplitude is of second order in the photon-hadron
interaction Hamiltonian.  The latter induces a shift in the pion mass
which is also formally of order $e^2$, and which is different in
finite and infinite-volume for massless photons (power-law corrections
are expected). A second minor complication arises due to the two
possible helicities of the final-state photons (either both positive
or both negative), since there are in principle two open elastic
channels. However, due to parity conservation, the two channels are
related by a phase difference and one can reduce the problem to a
single-channel problem.

The structure of this paper is as follows. The master equation
(\ref{eq:master}) relating the finite-volume matrix element to an
infinite-volume matrix element is derived for timelike photons in
section~2, where it is also brought into a form suited for lattice QCD
calculations.  In section~3, the pion decay amplitude is calculated in
two simple field theories coupled to massless photons, one being an
effective description for light quarks and the other for heavy quarks.
In section 4, the corresponding finite-volume matrix elements are
calculated directly, thus providing a check of the master
equation. Unless otherwise stated, we use the notation and conventions
of Peskin and Schroeder~\cite{Peskin:1995ev}, in particular the metric
convention $g_{\mu\nu}={\rm diag}(1,-1,-1,-1)$.

\section{General considerations}

We introduce the amplitude for the decay
$\pi_0\to\gamma^*\gamma^*$. In standard notation
(see for instance \cite{Jegerlehner:2009ry}~section~5.1), the amplitude is given by
$e^2$ times the QCD matrix element
\ba\la{eq:Apqmunu}
{\cal A}_{\mu\nu}(p,q) &=& 
i\int d^4 x \,e^{iq\cdot x}\,\<0 | {\rm T}\left\{j_\mu(x)j_\nu(0) \right\} | \pi_0(p)\>_{\rm rel}
\\ \la{eq:Fpq}
&=& \epsilon_{\mu\nu\alpha\beta}\, q^\alpha p^\beta \;{\cal F}(m_\pi^2,q^2,(p-q)^2),
\ea
where $j^\mu$ is the electromagnetic current of hadrons.
We have made explicit the use of the `relativistic' normalization of states,
\be
_{\rm rel}\<\pi_0(p')|\pi_0(p)\>_{\rm rel} = 2E_{\vec p} (2\pi)^3 \delta^{(3)}(\vec p-\vec p').
\ee
The quantity ${\cal F}$ is the off-shell $\pi_0\to\gamma\gamma$ form
factor. We will only consider the situation where the pion is
on-shell up to higher order corrections and (for simplicity) the case of two photons of same invariant mass,
\ba\la{eq:Apq}
{\cal A} \equiv \epsilon_{\sigma}^{\mu\,*}(q)\epsilon_{\sigma}^{\nu\,*}(p-q)
\int d^4 x \,e^{iq\cdot x}\,\<0 | {\rm T}\left\{j_\mu(x)j_\nu(0) \right\} | \pi_0(p)\>_{\rm rel},
\qquad \sigma=+,
\ea
with $p^2=M^2$ and $q^2=(p-q)^2$. We generically  denote by $\epsilon_{\sigma,\mu}(k)$ 
the polarization vector of the photon. Only the $\sigma=\pm1$ polarizations will play a role, 
because the longitudinal polarization $\sigma=0$ does not contribute to the amplitude.
The pole contribution to the Lorentz-invariant light-by-light scattering amplitude is then
\be\la{eq:iMpole}
i{\cal M}(\gamma_+\gamma_+\to\pi_0\to\gamma_+\gamma_+) = 
- e^4 {\cal A}^* \; \frac{i}{p^2-M^2} \; {\cal A}.
\ee
Except in the propagator, the four-momentum of the pion can be
considered to be on-shell.  Due to angular momentum conservation, the
two photons must have the same helicity.  Due to parity conservation,
there is a relative minus sign between the amplitudes for two
positive-helicity photons and the amplitude for two negative-helicity
photons in the final state.  This information is already encoded in
the parametrization (\ref{eq:Fpq}).

We will need the lowest term of the partial-wave expansion of the
light-by-light scattering matrix.  Due to the two possible helicity
states of the two photons in the final state, we are really dealing
with a (special case of the) two-channel problem. In the $s$-wave, the
partial wave expansion of the amplitude (\ref{eq:iMpole}) reads
(see appendix A)
\be\la{eq:pw}
{\cal M} = \frac{1}{2}\cdot 16\pi E\cdot t_0, 
\qquad t_0=\frac{1}{2ik}\left(e^{2i\delta_0}-1\right).
\ee

\subsection{Discrete energy eigenstates on the torus}

Our goal is to extract the decay rate of the neutral pion into two
photons, $\pi^0\to\gamma\gamma$, from stationary physics on the torus.
For the sake of the argument we will however consider massive vector
bosons $v$ with a finite mass $M_v\leq M/2$.  The QCD part of the
$\pi_0$ decay amplitude into two timelike photons is reinterpreted as
the QCD part of the $\pi_0$ decay amplitude into two massive, on-shell
vector bosons.

The linear size $L$  of the torus will be assumed to be in the range 
\be\la{eq:2piL}
\frac{2\pi}{L}\sim  \Lambda_{\overline{\rm MS}}.
\ee
In current state-of-the-art lattice QCD calculations, $L$ is typically 3 to 5 fm,
in which case $2\pi/L = 1240{\rm MeV}/L[{\rm fm}]$ is in the ballpark of (\ref{eq:2piL}).
Also, the pion is assumed to fit well into the box, so that the mass difference between
the finite-volume theory and  the infinite-volume theory can be neglected~\cite{Luscher:1985dn}.

We will restrict our attention to the center-of-mass system throughout this paper.
We work perturbatively in the electromagnetic coupling $e$. 
Initially we set $e=0$, so that the vector-boson states and the hadronic states
are decoupled. The spectrum of single-vector-boson states is
\be
\left\{ \omega_{\bk} \;\Big|\; \bk = \frac{2\pi}{L}\bn,\; \bn\in \mathbb{Z}^3 \right\},
\quad \qquad \omega_{\bk} = \sqrt{\bk^2+M_v^2},
\ee
and the spectrum of two-photon states with vanishing total momentum is
\be\la{eq:2GamSpec}
\left\{ 2\omega_{\bk} \;\Big|\; \bk = \frac{2\pi}{L}\bn,\; \bn\in \mathbb{Z}^3 \right\}.
\ee

The interaction Hamiltonian  reads
\be
V = +e \int \ud\bx\; j^\mu(x)\,A_\mu(x). 
\ee
The energy levels will be affected by the electromagnetic coupling.
Consider first non-degenerate states.
At first order in ordinary time-independent perturbation theory,
the non-degenerate energy levels will change by 
$ \Delta E_n^{(1)} = \< n | V | n \>$ . We normalize the finite-volume states 
to have unit norm, $\<n|n\>=1$.
However, the interaction Hamiltonian changes the vector-boson number by
exactly one unit, therefore the
matrix elements between states containing the same,
definite number of vector bosons vanish. At second order, 
\be\la{eq:DeltaE2}
\Delta E^{(2)}_n = \sum_{m\neq n} \frac{|V_{mn}|^2}{E_n^{(0)} - E_m^{(0)}}.
\ee
If the unperturbed state $|n\>$ is a non-degenerate, purely hadronic
state, then the states $|m\>$ that contribute non-trivially to the sum
in \eq(\ref{eq:DeltaE2}) are states that contain one vector boson and have
an arbitrary hadronic content. For a state $|n\>$ containing $N_n$
vector bosons, contributions come from states with $N_n\pm 1$ vector bosons.

Here we will be interested in the case where the box size is tuned so
that the neutral pion mass coincides with the energy of exactly one
two-vector-boson state with vanishing total momentum. As is well known, the
corrections to the spectrum are found by diagonalizing the perturbing
Hamiltonian in the degenerate subspace, which amounts to diagonalizing
a finite matrix $V_{nn'}$. In this case, the $2\times2$
matrix $V_{nn'}$ vanishes since there is no pair of states for which
the two states differ by a vector boson number of one.
To find the leading effect of e.m.\ interactions on the degenerate 
$|\pi_0\>$, $|vv\>$ states we therefore have to go to second 
order of degenerate perturbation theory. The recipe is then to diagonalize 
the matrix (\!\!\cite{LandauLif3}, paragraph 39) 
\be\la{eq:Wnn'}
W_{nn'}\equiv  V_{nn'} + \sum_{m} \frac{V_{nm} \, V_{mn'}}{E_n^{(0)}- E_m^{(0)} },
\ee
where the sum over $m$ extends over all states not in the degenerate
subspace.  Since $V_{nn'}$ vanishes, the matrix $W_{nn'}$ is
O($e^2$). We label the neutral pion at rest by `1' and the
two-vector-boson state by `2'. For instance, the matrix element $W_{22}$
corresponds to the transition of one vector boson to a hadronic state and
back, while the other one remains as a spectator,  a
vacuum polarization effect due to the hadrons\footnote{There is a
further potential effect, stemming from the fact that the
electromagnetic interaction breaks isospin symmetry, which is the
mixing of the $\pi_0$ with the $\eta$ meson and other, not necessarily
isovector states. Since the $\pi_0-\eta$ mixing starts at O($e^4$), we
can neglect it.}. The matrix element $W_{11}$ correspond to a mass 
correction for the pion due to the weakly coupled vector bosons. Its dependence on the volume 
is exponentially suppressed\footnote{For instance, for a non-relativistic bound state described by a 
wavefunction $\Psi(\vec r)$, it would be 
$\Delta M = -e^2\int d^3\vec r\; |\psi(\vec r)|^2 \frac{e^{-M_v r}}{4\pi r}$ in infinite volume and 
would only differ by an exponentially small amount on the torus, provided the bound state fits well into the box.}.
The matrix $W_{nn'}$ takes the form
\be
W = \left( \begin{array}{c@{\qquad}c}
\delta_L M &  W_{12} \\
     W_{12}^*       &  2\delta_L E_v
\end{array}
  \right)\,,
\ee
and the final energy eigenvalues are thus predicted to be, to O($e^2$),  
\be
E_\pm = M + \Delta E_\pm = M +\half{\delta_L M} + \delta_L E_v
\pm \sqrt{({\txts\frac{1}{2}}\delta_L M-\delta_L E_v)^2 + |W_{12}|^2}\,.
\ee

\subsection{Light-by-light scattering near the pole}

Hadronic light-by-light scattering is generically of order $e^4$,
however for energies a distance $e^2$ from the pion pole, the
scattering amplitude and the phase shift are of order $e^2$.  The
decay width of the $\pi_0$ is of order $e^4$ and therefore
negligible in the following considerations.  However, we must take
into account the O($e^2$)  shifts in the $\pi_0$
mass and the photon mass.

We treat $\delta_0$ as a function of the vector boson momentum. 
For a small phase shift, \eq(\ref{eq:iMpole}) and (\ref{eq:pw}) yield
\be\la{eq:d0pm}
\delta_0(k_\pm) = - \frac{ke^4 |{\cal A}|^2}{16\pi M^2(\Delta E_{\pm}-\delta_\infty M)}.
\ee

\subsection{Relation of the discrete energy levels to the $\pi_0\to vv$ amplitude}

In the  finite-volume theory, we  work in the basis formed by the state
\be\la{eq:GG-}
|(vv)_-\> = \frac{1}{2\sqrt{\nu_n}}\sum_{\vec k\in\Omega_n} \sum_{\sigma=\pm1} \sigma ~ 
a^\dagger_{\vec k\sigma} a^\dagger_{-\vec k\sigma} |0\>
\ee
and the state $|(vv)_+\>$ for which the relative sign between the two
helicity terms is positive. The set of momenta appearing in
\eq(\ref{eq:GG-}) is 
\[
\Omega_n=\{\vec k=2\pi\vec z/L\;|\; \vec z\in\mathbb{Z}^3,\;\vec z^2=n\},
\]
where we restrict ourselves to a value of $n\leq 6$ in order to avoid
multiple degeneracies, and $\nu_n=|\Omega_n|$ is the cardinality of
that set.  Only $|(vv)_-\>$ couples to the $\pi_0$, so that we can
ignore the state $|(vv)_+\>$ to the order $e^2$ we are working at. The
discrete set of momenta $\Gamma$ comprises all the images under the
cubic group of one particular allowed momentum on the torus.

Using L\"uscher's finite-volume formalism~\cite{Luscher:1986pf,Luscher:1990ux}, a
change in the effective momentum $k$ defined via the two-vector-boson
spectrum, $E = 2\sqrt{M_v^2+k^2}$, from the free-field values (\ref{eq:2GamSpec})
corresponds to an $s$-wave scattering phase~\cite{Lellouch:2000pv,Meyer:2011um}
\be
\Delta\delta_0(k) = - \left( q\phi'(q) + k\frac{\partial \delta_0(k)}{\partial k}\right) 
\frac{\Delta k}{k} ,\qquad q\equiv \frac{kL}{2\pi}
\ee
with $\phi$ a tabulated kinematic function.
Here $\delta_0(k)$ vanishes before the vector bosons are coupled to hadrons.
Note that the electromagnetic shift in the $\pi_0$ pole is contained in $\Delta\delta_0$.
We use the derivative of  $\phi$ at a point where $q=|\vec z|$ 
for some vector $\vec z\in\mathbb{Z}^3$,
$
q\phi'(q) = (2\pi)^2 \frac{q^3}{\nu_n}
$.
One thus finds
\be\la{eq:d0FV}
\delta_0(k_\pm) = -\frac{L^3 k^2}{2\pi\nu_n}\,\Delta k_\pm .
\ee
Now the change in the effective momentum of the two vector bosons is related to the spectrum via
\be\la{eq:kin}
\Delta k_\pm = \frac{M}{4k} \left( \Delta E_{\pm} -2\delta_L E_v\right).
\ee
Combining \eq(\ref{eq:d0pm}), (\ref{eq:d0FV}) and (\ref{eq:kin}), we obtain the master relation
\be\label{eq:master}
|{\cal A}|^2 =  \frac{2L^3M^3}{e^4\nu_n} \cdot (\Delta E_\pm -\delta_\infty M)(\Delta E_\pm -2\delta_L E_v)
 = \frac{2L^3M^3}{e^4\nu_n} \cdot  |W_{12}|^2.
\ee
We have used the fact that $\delta_L M = \delta_\infty M$ up to exponential corrections, and neglected the latter.

\subsection{Explicit expression for the matrix element $W_{12}$}

Using the spectral representation, one verifies that $W_{12}$, defined in \eq(\ref{eq:Wnn'}), 
is given by 
\be
W_{12} = - \frac{1}{2} \int_{-\infty}^\infty dt\; \<(vv)_-| {\rm T}\{V(t)V(0)\} | B\>,
\ee
as long as there is no contributing intermediate hadronic state with energy $E_h\leq M/2$.
The Euclidean time evolution, $V(t) = e^{Ht} V(0) e^{-Ht}$, is dictated by the full QCD+mQED Hamiltonian 
$H$.
The Wick contractions of the vector bosons can be carried out straightforwardly by noting that
in equation (\ref{eq:Wnn'}), the intermediate states contributing are direct products 
of an arbitrary hadronic state with a one-vector-boson state and by using equation (\ref{eq:GG-})
and the plane-wave expansion of the gauge field. One then reduces $W_{12}$ to a pure QCD matrix element,
\be\la{eq:W12simple}
W_{12} = -{e^2\sqrt{\nu_n}}  \epsilon_{+\mu}^*(-\vec k) \epsilon_{+\nu}^*(\vec k)
 \int d^4x\; \frac{e^{\omega_k |x_0|+i\vec k\cdot\vec x}}{2\omega_k}\,\<0|{\rm T}\{j^\mu(x) j^\nu(0)\}|B\>.
\ee
Taking into account the relativistic normalization of states ($
\sqrt{2ML^3} |B\> \to |B\>_{\rm rel.} $) and inserting
(\ref{eq:W12simple}) into (\ref{eq:master}) shows that, possibly up to an overall phase $\varphi$, the
$\pi_0\to vv$ amplitude is given by the analytic continuation of \eq(\ref{eq:Apq}),
\be\la{eq:Afinal}
{\cal A} = e^{i\varphi} \epsilon_{+\mu}^*(-\vec k) \epsilon_{+\nu}^*(\vec k)
 \int d^4x\; e^{\omega_k |x_0|+i\vec k\cdot\vec x}\,\<0|{\rm T}\{j^\mu(x) j^\nu(0)\}|B\>_{\rm rel}.
\ee
This expression is suitable for an implementation in lattice
QCD~\cite{Ji:2001wha}, see~\cite{Feng:2012ck,Lin:2013im,Feng:2012an}
for recent calculations.

\section{The pion decay amplitude in two simple field theories}

In the previous section, relation (\ref{eq:master}) was derived for
massive vector bosons and then reduced to a pure QCD matrix element,
\eq(\ref{eq:Afinal}). In the latter equation, the only memory of the
dispersion relation of the vector bosons is the relation between
$\omega_k$ and the spatial momentum $\vec k$.  Since the vector bosons
merely appear as external legs in the amplitude, we expect that the
volume corrections affecting the calculation of the amplitude ${\cal
  A}$ in finite volume based on \eq(\ref{eq:Afinal}) are exponentially
small in the box size, \emph{even for the case of real photons} where
$\omega_k=|\vec k|$.

In order to check this expectation, in this section we calculate, in two tractable
field theories, the amplitude ${\cal A}$ for on-shell photons in the
final state. We then use relation (\ref{eq:master}) to predict the
finite-volume quantity $|W_{12}|^2$.  In the next section, the same
quantity will be computed directly in finite volume as a check of
\eq(\ref{eq:master}).

\subsection{Theory I: chiral Lagrangian}

In chiral effective theory the decay of a neutral pion of mass $M$
into two photons is driven by the effective Lagrangian (see for
instance~\cite{Weinberg:1996kr}, chapter 22)
\be
{\cal L}_a = g\;\pi_0 \;\epsilon^{\mu\nu\rho\sigma} F_{\mu\nu}F_{\rho\sigma}.
\ee
In the chiral limit the Abelian chiral anomaly~\cite{Adler:1969gk,Bell:1969ts} 
predicts the value of the coupling~\cite{Adler:1969er},
\be
g = \frac{N_c e^2}{48\pi^2 F_\pi}
\ee
with $F_\pi\simeq 184{\rm MeV}$ the pion decay constant and $N_c$ the number of colors,
however we will not assume any particular value for $g$.
In this theory the module of the invariant amplitudes for $\pi_0$ decay then amounts to
\be
|{\cal M}(B\to \gamma_+\gamma_+)|^2 = |{\cal M}(B\to \gamma_-\gamma_-)|^2 = e^4 |A|^2 = 16g^2 M^4,
\ee
which leads to a full $\gamma\gamma$ width of $\Gamma = M^3 g^2/\pi$.

In order to verify (\ref{eq:master}), we thus have to show by a direct calculation 
that the module of the finite-volume matrix element $W_{12}$ is given by 
\be\la{eq:target0}
|W_{12}|^2 = \frac{8\nu_n M g^2}{L^3}\;.
\ee

\subsection{Theory II: a non-relativistic bound state of two massive fermions}

Consider first the amplitude for the fermion-pair annihilation  $\bar f f \to \gamma\gamma$.
The fermions have a mass $m$.
The initial momenta of the fermions are $p$ and $p'$ and the final momenta of the photons are $k$ and $k'$.
There are two diagrams at treelevel, yielding
\be
{\cal M} = \frac{e^2}{2} \epsilon_\mu(k)^* \epsilon_\nu(k')^*\;
\bar v(p') \left[\gamma^\nu \frac{(2p^\mu-\slash\!\!\!{k}\gamma^\mu)}{p\cdot k} 
         + \frac{\gamma^\mu (2p^\nu-\slash\!\!\!{k}'\gamma^\nu)}{p\cdot k'}\right] u(p)
\ee
We have used $p^2=m^2$ and $k^2=0$ and the spinor identities 
\be
(\slash\!\!\!p+m)\gamma^\nu u(p) = 2p^\nu u(p), \qquad 
\overline u(p)\gamma^\nu (\slash\!\!\!p+m) = \overline u(p) 2p^\nu.
\ee

Now we choose the center-of-mass frame and consider the limit of non-relativistic fermions. 
Using standard spinor technology one obtains (for an initial state with one spin-up fermion 
and one spin-down fermion)
\be
{\cal M} = \frac{2i e^2}{m}\, (\vec k\times \vec\epsilon_\sigma(\vec k)^*)\cdot \vec\epsilon_{\sigma'}(-\vec k)^*.
\ee
For instance, if $\vec k= |\vec k| \vec e_3$ and $\vec\epsilon_\sigma(\vec k) = \frac{1}{\sqrt{2}}(1,i\sigma,0)$,
$\vec\epsilon_{\sigma'}(-\vec k) = \frac{1}{\sqrt{2}}(1,-i\sigma',0)$, then
\be
i(\vec e_3\times \vec\epsilon_\sigma(k)^*)\cdot \vec\epsilon_{\sigma'}(k')^*  = -\sigma\,\delta_{\sigma\sigma'}.  
\ee

Following (\!\!\cite{Peskin:1995ev}, chapter 5), we describe a
pseudoscalar bound state as\footnote{ The first factor in (\ref{eq:B})
  converts the state from a non-relativistic normalization to a
  relativistic one and we have defined
\be
\Big|\,\vec p\uparrow,\;-\vec p\downarrow\Big\> = a_{\vec p(+)}^{F\,\dagger} a_{-\vec p(-)}^{{\bar F}\,\dagger} \;\Big|\, 0\Big\>,
\qquad \quad
\Big|\,\vec p\downarrow,\;-\vec p\uparrow\Big\> = a_{\vec p(-)}^{F\,\dagger} a_{-\vec p(+)}^{\bar F\,\dagger} \;\Big|\, 0\Big\>,
\ee
where $a^F$ ($a^{\bar F}$) is the annihilation operator of the particle 
(antiparticle). The relative sign
  between the two terms in (\ref{eq:B}) can be checked by calculating
  the (non-vanishing) amplitude $\<0|\bar\psi\gamma_5\psi | B\>$.}
\be\label{eq:B}
|B\> = \sqrt{2M} \int \frac{d^3\vec p}{(2\pi)^3}\; 
\tilde \Psi(\vec p) \frac{1}{\sqrt{2m}}\frac{1}{\sqrt{2m}} \frac{1}{\sqrt{2}}
\left( |\vec p\uparrow,\;-\vec p\downarrow\>_{\rm rel} + |\vec p\downarrow,\;-\vec p\uparrow\>_{\rm rel}\right).
\ee
The amplitude for $B\to\gamma\gamma$ is then
\ba
{\cal M}(B\to\gamma_\sigma\gamma_{\sigma'}) &=& \sqrt{2M}\int \frac{d^3\vec p}{(2\pi)^3}\;
  \frac{\tilde\Psi(\vec p)}{{2m}} \cdot 
\left( {\cal M}(\vec p\uparrow,\,-\vec p\downarrow\to \gamma\gamma)
+  {\cal M}(\vec p\downarrow,\,-\vec p\uparrow\to \gamma\gamma)
\right)
\nn
&=& -\sigma \delta_{\sigma\sigma'}\;\frac{2\sqrt{2}e^2}{\sqrt{m}} \;\Psi(\vec x=0).
\la{eq:MggNR}
\ea
As a check, the decay rate of the bound state of mass $M$ into two photons now reads 
\ba
\Gamma &=& \frac{1}{2M}\; \int\frac{d^3\vec k}{(2\pi)^3 2|\vec k|}\; \int\frac{d^3\vec k'}{(2\pi)^3 2|\vec k'|}\; 
\, (2\pi)^4\, \delta^{(4)}(p+p'-k-k')\, |{\cal M}(B\to\gamma\gamma)|^2
\nn
&=& \frac{e^4}{4\pi m^2}    |\Psi(\vec x=0)|^2\,.
\ea
In $\Gamma$, we have implicitly included a factor 2 stemming from the sum over final polarization
states of the photons, $\sum_{\sigma}\sum_{\sigma'}\delta_{\sigma\sigma'} = 2$, but also a factor
$\frac{1}{2!}$ to remove indistinguishable final-state configurations of the photons.
For muonium in the ground state, we have
$\Psi(\vec x=0) = {(\mu^{3/2} e^3)}/{(8\pi^2)}$,
with $\mu=m/2$ the reduced mass, and the width becomes the well-known result
$\Gamma % = \frac{e^4}{4\pi m^2}   m^3/8  e^6 / (64\pi^4)    = e^10 m /(2^11 \pi^5) = 
 = {\alpha^5 m}/{2}$.

By comparison with \eq(\ref{eq:Apq}), the result (\ref{eq:MggNR}) implies in particular 
\be
|A|^2 = \frac{8}{m}|\Psi(\vec x=0)|^2.
\ee
In order to verify (\ref{eq:master}), we thus have to show by a direct calculation 
that the module of the finite-volume matrix element $W_{12}$ is given by 
\be\la{eq:target}
|W_{12}|^2 = \frac{e^4\nu_n}{2L^3 m^4} |\Psi(\vec x=0)|^2.
\ee

\section{Analytic calculation of finite-volume matrix elements}

In order to check the master formula (\ref{eq:master}), we need to
determine by a direct calculation the norm of the matrix element
$W_{12}$. We work in the Euclidean theory and adopt the corresponding
notation. In particular, the Euclidean Dirac matrices are related to
the Minkowski space matrices by $\gamma^{_E}_0=\gamma^0$ and
$\gamma_k^{_E} = -i\gamma^k$, $\gamma_5^{_E}=\gamma_5$, but we will
not display the superscript `E' explicitly.

\subsection*{General considerations and the spectral representation}

In order to compute $W_{12}$, we consider the gauge-invariant correlation function
\be
G(t) =  \left\<  {\cal P}(t)  {\cal O}_5(0)\right\>.
\ee
A gauge-invariant choice of operators would be 
\ba
{\cal P}(t) &=& \frac{1}{L^{3/2}}\int d^3\vec x\; \epsilon_{\mu\nu\rho\sigma} F_{\mu\nu}(t,\vec x) F_{\rho\sigma}(t,\vec x)
\\
{\cal O}_5(t) &=& \frac{1}{L^{3/2}}\int d^3\vec x\; \bar\psi(t,\vec x)\gamma_5 \psi(t,\vec x).
\ea
In order to simplify the perturbative calculation  we will however make a slightly different choice in the next section.
We will compute $G(t)$ perturbatively in the electromagnetic interaction 
at order $e^2$. The first non-trivial contribution to $G(t)$ occurs at order $e^2$, 
\ba
G(t) &=&  G_0(t) + e^2 G_2(t) + e^4 G_4(t) + \dots,
\\
G_2(t) &=& {\txts\frac{1}{2!}}\left\< {\cal P}(t) 
\left({\txts\int} d^4z\,{\cal V}(z)\right)^2  {\cal O}_5(0) \right\>_0,
\\
{\cal V}(x) &=& -ie A_\mu(x) \bar\psi(x)\gamma^\mu\psi(x),
\\
V(t) &=& \int d^3\vec x\; {\cal V}(t,\vec x).
\ea
In these equations the (Euclidean) time dependence of the operators is
dictated by the non-QED part of the Hamiltonian.

On one hand, the correlation function can be interpreted in terms of the spectral representation.
In order to isolate the desired matrix element, we notice that the contribution of interest comes
about when the interaction Hamiltonian $\int d^3\vec x\; {\cal V}$ is inserted between time
$0$ and $t$,
\ba
&& G_2(t)=  \sum_{n,n'} \Big\<0\Big| {\cal P}(t)\Big|n\Big\>\;
\int_0^t ds\;  e^{-E_n(t-s)} \int_0^s ds'
\\ && \qquad \qquad 
 \sum_l   \< n| V |l\> \; e^{-E_l(s-s')} \<l| V  |n'\>\,e^{-E_{n'}s'} 
\;   \Big\< n' \Big| {\cal O}_5(0) \Big| 0 \Big\> +\dots
\nonumber
\ea
We assume that the torus size has been tuned so that 
the states $|n\>=|\gamma\gamma\>$ and $|n'\> = |B\>$ are degenerate with an energy 
$E = 2|\vec k|$. 
Carrying out the integrals over $s$ and $s'$, their contribution to $G_2(t)$
will have the following dependence on $t$,
\be
G_2(t) =  t e^{-Et} 
({\cal P}|\gamma\gamma) (B|{\cal O}_5) 
\left(\sum_l \frac{\<\gamma\gamma|V|l\> \;\<l|V|B\>}{E_l-E}\right) + \dots
\ee
where for brevity we have defined
\be
({\cal P}|\gamma\gamma)\equiv   \Big\<0\Big|{\cal P}(0)\Big|\gamma\gamma\Big\>,
\qquad \qquad 
(B|{\cal O}_5) \equiv  \Big\< B \Big| {\cal O}_5(0) \Big| 0 \Big\> .
\ee
By comparing to \eq(\ref{eq:Wnn'}), we see that we can extract the matrix element $W_{12}$ from
\be\la{eq:expform}
G_2(t) = - {W_{12}}\cdot ({\cal P}|\gamma\gamma)\cdot (B|{\cal O}_5)
  \, t\, e^{-Et} + \dots
\ee

\subsubsection*{The overlap factors}

We use creation and annihilation operators that are unit-normalized,
$[a_{\vec p \underline{s}},a_{\vec p'\underline{s}'}^\dagger]_\pm = \delta_{\underline{ss}'} \delta_{\vec p\vec p'}$
where the plus sign is for bosons and the minus sign for fermions and $\underline{s}$ represents all discrete
indices characterizing the particles.
The unit-normalized energy eigenstates we are interested in are the following.
The two-photon state is defined in \eq(\ref{eq:GG-}).
For Theory I and II respectively, the `pseudoscalar meson' states are
\ba
|B\> &=& a_{\vec p=0}^\dagger \,|0\>,
\\
|B\> &=&  \frac{1}{\sqrt{2L^3}} \sum_{\vec p}
\tilde \Psi(\vec p) \Big(a_{\vec p(+)}^{F\,\dagger} a_{-\vec p(-)}^{{\bar F}\,\dagger} 
             + a_{\vec p(-)}^{F\,\dagger} a_{-\vec p(+)}^{\bar F\,\dagger}  \Big) \;\Big|0\Big\>
\ea
We will use the following spatial Fourier representation and normalization for the wave function,
\be
\Psi(\vec x) = \frac{1}{L^3} \sum_{\vec p} e^{i\vec p\cdot\vec x} \tilde \Psi(\vec p),
\qquad \int d^3\vec x\; |\Psi(\vec x)|^2 = \frac{1}{L^3} \sum_{\vec p} |\tilde\Psi(\vec p)|^2 = 1.
\ee

Our interpolating operator for the two-photon state will be
\be
{\cal P} = A_i(-\vec k) A_j(\vec k),\qquad \qquad {\rm with}\quad  i=1,~ j=2,~ \vec k = |\vec k| \vec e_3.
\ee
Using the plane-wave expansion of the gauge fields,
\be\la{eq:AmuPlW}
A_\mu(\vec x) = \sum_{\vec k,\sigma} \frac{1}{\sqrt{2\omega_k L^3}} 
\left( a_{\vec k\sigma}\, \epsilon_{\sigma\mu}(\vec k) \, e^{i\vec k\cdot\vec x}
+ a_{\vec k\sigma}^\dagger \, \epsilon^*_{\sigma\mu}(\vec k)\, e^{-i\vec k\cdot\vec x}\right),
\ee
one finds 
\be\la{eq:Pgg}
\big({\cal P}\big|(\gamma\gamma)_-\big) = \frac{iL^3}{2\omega_k \sqrt{\nu_n}}.
\ee
for the matrix element.
The matter fields have the expansion
\ba
\pi_0(\vec x) &=& \sum_{\vec p} \frac{1}{\sqrt{2E_{\vec p}L^3}}
\left(a^{\pi_0}_{\vec p}\,e^{i\vec p\cdot\vec x} + a_{\vec p}^{\pi_0\,\dagger} \, e^{-i\vec p\cdot\vec x}\right),
\\
\psi(\vec x) &=& \sum_{\vec p}\frac{1}{\sqrt{2E_p L^3}} \sum_s\left(
a^F_{\vec ps}\, u^s(\vec p) e^{i\vec p\cdot\vec x} + a_{\vec ps}^{\bar F\,\dagger}\, v^s(\vec p) e^{-i\vec p\cdot\vec x}\right),
\\
\bar\psi(\vec x) &=& \sum_{\vec p}\frac{1}{\sqrt{2E_p L^3}} \sum_s\left(
a_{\vec ps}^{\bar F}\, \bar v^s(\vec p)\, e^{i\vec p\cdot\vec x} + a_{\vec ps}^{F\,\dagger} \,\bar u^s(\vec p)\, e^{-i\vec p\cdot\vec x}
\right)
\ea
with the normalization $\bar u^r(\vec p) u^s(\vec p) = 2m \delta^{rs}$ and 
$\bar v^r(\vec p) v^s(\vec p) = -2m \delta^{rs}$.
In theory I, the interpolating operator is simply the pion field, ${\cal O}_5=\pi_0$,
and the overlap is
\be
\la{eq:BO5o}
(B|{\cal O}_5) = \left(\frac{L^3}{2M}\right)^{1/2}\,;
\ee
in theory II, the interpolating operator is chosen to be 
${\cal O}_5=-\bar\psi(\vec p)\gamma_5\psi(-\vec p)$ and the overlap is
\be\la{eq:BO5}
(B|{\cal O}_5) = -\sqrt{2L^3}\,\tilde\Psi^*(-\vec p).
\ee

\subsubsection*{Perturbation theory in the time-momentum representation}

On the other hand, $G(t)$ can be computed using the standard techniques of
Feynman diagrams, using the time-momentum representation of the
propagators.  The fields are spatially Fourier transformed according
to
\ba
A_\mu(t,\vec k) = \int d^3\vec x\; A_\mu(t,\vec x)\; e^{-i\vec k\cdot\vec x},
&\quad& \pi_0(t,\vec k) = \int d^3\vec x\; \pi_0(t,\vec x)\; e^{-i\vec k\cdot\vec x},\qquad 
\\
\psi(t,\vec k) = \int d^3\vec x\; \psi(t,\vec x)\; e^{-i\vec k\cdot\vec x},
&\quad& \bar\psi(t,\vec k) = \int d^3\vec x\; \bar\psi(t,\vec x)\; e^{-i\vec k\cdot\vec x}.\qquad 
\ea
In this representation, the photon propagator in Feynman gauge reads
\be
\<A_\mu(t,\vec k)\, A_\nu(t',-\vec k')\>
=\delta_{\mu\nu} L^3 \delta_{\vec k\vec k'} \;\frac{e^{-\omega_k |t-t'|}}{2\omega_k}\,,
\ee
and the matter fields have the propagators
\ba
\<\pi_0(t,\vec p)\, \pi_0(t',-\vec k') \> &=&
L^3 \delta_{\vec p\vec p'} \;\frac{e^{-E_{\vec p} |t-t'|}}{2E_{\vec p}}\,,
\\
\<\psi(t,\vec p)\,\bar\psi(t',-\vec p')\>
&=& L^3 \delta_{\vec p\vec p'}\Big(E_p\,{\rm sgn}(t-t')\,\gamma_0 -i\vec p\cdot\vec\gamma + m\Big)\; \frac{e^{-E_{p}|t-t'|}}{2E_p}\;.
\qquad 
\ea
The fermion-photon vertex is $-ie\gamma_\mu$ ($\mu$ corresponds to the Lorentz index of the photon and $e$ 
is the charge of the fermion).

\subsection{Theory I: finite-volume calculation}

At first, we will not make the assumption that the two-photon state is necessarily 
degenerate with the pseudoscalar matter state.
The correlation function at order $e^2$ with all fields written in the time-momentum 
representation reads
\ba
G_2(t) &=& ig \epsilon_{\mu\nu\rho\sigma} \Big\< A_1(t,-\vec k) A_2(t,\vec k)
\\ && \quad \int ds \frac{1}{L^3}\sum_{\vec q}
\pi_0(s,\vec q) \frac{1}{L^3} \sum_{\vec \ell} F_{\mu\nu}(s,\vec l) F_{\rho\sigma}(s,-(\vec l+\vec q))\; \pi_0(0,\vec p=0)\Big\>.\qquad 
\nonumber
\ea
After all Wick contractions have been performed, one arrives at 
\be
G_2(t) = -\frac{2gk_3L^3}{M\omega_k}\int_{-\infty}^\infty ds \; {\rm sgn}(t-s)\, e^{-(M|s|+2\omega_k|t-s|)}.
\ee
Now carrying out the integral over $s$ yields
\ba\la{eq:poleL}
\int_{-\infty}^\infty ds \; {\rm sgn}(t-s)\, e^{-(M|s|+2\omega_k|t-s|)}
= \frac{e^{-2\omega_k t}- e^{-Mt}}{2\omega_k+M} + \frac{e^{-Mt}- e^{-2\omega_k t}}{2\omega_k-M}\;,
\ea
where the first term comes from the region $s<0$ and $s>t$ and the second from $0<s<t$.
If we now take the limit $\omega_k\to M/2$, the first term vanishes and the second becomes
$t e^{-Mt}$. Thus the correlation function becomes, for $k_3 = M/2$,
\be
G_2(t)= -\frac{2gL^3}{M} t \, e^{-Mt}.
\ee
Comparing it with \eq(\ref{eq:expform}) and dividing out the overlap factors
(\ref{eq:Pgg}) and (\ref{eq:BO5o}), we indeed arrive at \eq(\ref{eq:target0}), 
thus confirming the master relation (\ref{eq:master}).

\subsection{Theory II: finite-volume calculation}

The correlation function at order $e^2$ with all fields written in the time-momentum 
representation reads
\ba
&& G_2(t) =  -\frac{e^2}{2} \Big\<A_i(t,-\vec k) \, A_j(t,\vec k)
\\ && \quad 
\int_{-\infty}^\infty ds \, \frac{1}{L^3}\sum_{\vec \ell'}\frac{1}{L^3}\sum_{\vec \ell}\, A_\mu(s,\vec\ell') 
\int_{-\infty}^\infty ds' \frac{1}{L^3}\sum_{\vec q'} \frac{1}{L^3}\sum_{\vec q} A_\nu(s,\vec q') \;(\gamma_5)_{\beta\alpha}
\nn && \quad 
\bar\psi(s,\vec \ell)\gamma^\mu\psi(s,-(\vec \ell'+\vec \ell))\; 
\bar\psi(s',\vec q)\gamma^\nu\psi(s',-(\vec q'+\vec q))\; 
\psi_\alpha(0,-\vec p) \bar\psi_\beta(0,\vec p)\Big\>\; . \qquad 
\nonumber
\ea
The Wick contractions of the gauge fields lead to
\ba
&& G_2(t) = -\frac{e^2}{2} \int_{-\infty}^\infty ds  \int_{-\infty}^\infty ds'
\frac{e^{-\omega_{\vec k}|t-s|}}{2\omega_{\vec k}}  \frac{e^{-\omega_{\vec k}|t-s'|}}{2\omega_{\vec k}}
\frac{1}{L^3}\sum_{\vec \ell}\frac{1}{L^3}\sum_{\vec q} \; (\gamma_5)_{\beta\alpha} \qquad \qquad 
\\ && \quad 
\Big\<\bar\psi(s,\vec \ell)\gamma^i\psi(s,-(\vec \ell+\vec k))\;
\bar\psi(s',\vec q)\gamma^j\psi(s',\vec k-\vec q)\;
\psi_\alpha(0,-\vec p) \bar\psi_\beta(0,\vec p)\Big\>
\nn && \qquad  + (i\leftrightarrow j,~ \vec k\to -\vec k).
\nonumber
\ea
The fermion propagator joining the two insertions of the vector current is treated as a free propagator,
\ba
&& G_2(t) =  -{e^2} \int_{-\infty}^\infty ds\int_{-\infty}^\infty ds' 
\frac{e^{-\omega_{\vec k}|t-s|}}{2\omega_{\vec k}}  \frac{e^{-\omega_{\vec k}|t-s'|}}{2\omega_{\vec k}} 
\gamma^i_{\gamma\delta} \gamma^j_{\gamma'\delta'}\, (\gamma_5)_{\beta\alpha}
\\ && \quad 
\frac{1}{L^3}\sum_{\vec q}  
\frac{e^{-E_{\vec q+\vec k}|s-s'|}}{2E_{\vec q+\vec k}} 
\Big(E_{\vec q+\vec k}{\rm\,sgn}(s-s')\gamma_0 + i\vec \gamma\cdot(\vec q+\vec k) + m\Big)_{\delta\gamma'}
\nn && \qquad \qquad \qquad \qquad 
\Big<\bar\psi_\gamma(s,\vec q) \psi_{\delta'}(s',-\vec q)\, \psi_\alpha(0,-\vec p) \bar\psi_\beta(0,\vec p)\Big\>
\nn && \quad 
+ (i\leftrightarrow j,~ \vec k\to -\vec k).
\ea
We can now use expression (\ref{eq:f4pt}) for the fermion four-point function that takes into account the
presence of a bound state in the spectrum (see appendix B) and obtain
\ba \la{eq:G2bc}
&& G_2(t) = {2e^2}  \tilde\Psi(-\vec p)^* 
 \left(\bar u^{(+)}(-\vec p)\gamma_5 v^{(-)}(\vec p)  +  \bar u^{(-)}(-\vec p)\gamma_5 v^{(+)}(\vec p)\right)
\\ && \quad \int_{-\infty}^\infty ds\int_{-\infty}^s ds' 
\frac{e^{-\omega_{\vec k}|t-s|}}{2\omega_{\vec k}}  \frac{e^{-\omega_{\vec k}|t-s'|}}{2\omega_{\vec k}} \;\frac{e^{- E_B s'}}{2\cdot2E_p}  
\sum_{\vec q} \frac{e^{-E_{\vec q+\vec k}(s-s')}}{2E_{\vec q+\vec k}} \, \frac{e^{-E_{\vec q}(s-s')}}{2E_{\vec q}}\,  \tilde \Psi(-\vec q) 
\nn && \quad \sum_{r=\pm}  \bar v^{r}(\vec q) \gamma^i
\Big(E_{\vec q+\vec k}{\rm\,sgn}(s-s')\gamma_0 + i\vec \gamma\cdot(\vec q+\vec k) + m\Big) \gamma^j\, u^{-r}(-\vec q)
\nn && \quad 
+\; (i\leftrightarrow j,~ \vec k\to -\vec k) ~ + \dots
\nonumber
\ea
The dots stand for terms which do not contribute to the contribution proportional to $t\; e^{-Et}$ 
that we are interested in. 
Now since the wavefunction $\tilde \Psi(\vec q)$ is concentrated at momenta
much less than $m$ in norm, we can approximate all the factors at
leading order by evaluating them at $\vec q=0$. After this operation, the first two lines 
in \eq(\ref{eq:G2bc}) are an even function of $\vec k$, hence 
we can evaluate the spinor matrix element
where we now set $i=1$, $j=2$ and $\vec k = m \vec e_3$,
\ba
&&  \sum_{r=\pm}  \bar v^{r}(\vec 0) \gamma^i
\Big(E_{\vec k}{\rm\,sgn}(s-s')\gamma_0 + i\vec \gamma\cdot\vec k + m\Big) \gamma^j\, u^{-r}(\vec 0)
+ (i\leftrightarrow j,~ \vec k\to -\vec k) \qquad \qquad 
\\ && \qquad = -2im \sum_{r=\pm} v^{r}(\vec 0)^\dagger\gamma_5 u^{-r}(\vec 0) = 8im^2\;,
\nn
&& \bar u^{(+)}(\vec 0)\gamma_5 v^{(-)}(\vec 0)  +  \bar u^{(-)}(\vec 0)\gamma_5 v^{(+)}(\vec 0)= -4m.
\ea
Now the integrals over $s$ and $s'$ can be carried out, yielding (for $i=1$ and $j=2$)
\be
G_2(t) = - i\Psi(\vec 0)\,\tilde\Psi(-\vec p)^*\,\frac{e^2 L^3\, t\, e^{-2mt}}{2m^3} + \dots
\ee
We assume that the wavefunction $\Psi$ of the bound state on the torus
only differs by a negligible amount from the infinite volume
wavefunction. For $L$ sufficiently large this assumption is justified
by the general analysis~\cite{Luscher:1985dn}.  Comparing with
\eq(\ref{eq:expform}) and dividing by the overlap factors
(\ref{eq:BO5}) and (\ref{eq:Pgg}), we arrive at the expected formula
(\ref{eq:target}) and have thus confirmed the master relation
(\ref{eq:master}).

\section{Final remarks}

The approach followed here relies on the relation between the discrete
two-particle spectrum on the torus and the $S$-matrix of the
corresponding infinite-volume
theory~\cite{Luscher:1986pf,Luscher:1990ux}. Since this relation is
only guaranteed to hold up to exponential corrections in the volume
for massive field theories, we were led to consider massive vector
bosons. However, in the final QCD matrix element (\ref{eq:Afinal}),
the dispersion relation of the vector bosons only appears to play a
secondary role.  It would therefore be desirable to have a more direct
approach to the question of finite-size corrections of the
$\pi_0\to\gamma\gamma$ decay amplitude, presumably using some of the
mathematical techniques developed
in~\cite{Luscher:1986pf,Kim:2005gf,Hansen:2012tf}.

In isospin-symmetric QCD, the $\eta$ is perfectly stable, and it would
be interesting to study the case $\eta\to\gamma\gamma$ along the same
lines we followed here.  One must however take into account that at
$\sqrt{s}=M_\eta$, $\gamma\gamma\to 3\pi$ is kinematically allowed and
of the same order in the fine structure constant as the $\eta$ pole
contribution to light-by-light scattering. Unless one makes use of the
specific dynamics of pions~\cite{Gasser:1984pr}, one is then dealing
with a case involving an open inelastic channel. The theoretical
framework to study such problems is under
construction~\cite{Polejaeva:2012ut,Briceno:2012rv}.

\section*{Acknowledgments}
I thank Stefan Scherer for encouragement and interesting discussions.
This work was supported by the \emph{Center for Computational Sciences in Mainz}.

\appendix

\section{Reduction to a one-channel problem}

We consider the elastic  scattering of two vector bosons with same helicity.
In the final state, the vector bosons can have either both positive, or both negative
helicity, and there are thus two open channels.
We follow the notation of Sharpe and Hansen~\cite{Hansen:2012tf}.

The S-matrix is written as 
\be
i({\cal M})_{\rm B} = K ((S)_{\rm B}-1),\qquad K=\frac{8\pi E}{k}.
\ee
In the basis ${B}=(|v_+v_+\>,|v_-v_-\>)$, 
we saw that the scattering amplitude takes the form
\be\la{eq:MbasisDP}
i({\cal M})_{\rm B} = i\left(\begin{array}{cc}
{\cal M} & -{\cal M} \\
-{\cal M}  & {\cal M} 
\end{array}\right).
\ee
On the other hand, in order to relate ${\cal M}$ to a phase shift, we realize
 that in the basis $B'=(|(vv)_+\>,|(vv)_-\>)$, the S-matrix is given by
\be
(S)_{\rm B'} = {\rm diag}(e^{2i\delta_0},0).
\ee
Performing the change of basis,
\be
i({\cal M})_{\rm B} = {K} \left[\frac{1}{2}\left(\begin{array}{cc} 1 & 1 \\ -1 & 1 \end{array}\right)
\left(\begin{array}{cc} e^{2i\delta_0} & 0 \\ 0 & 1 \end{array}\right)
\left(\begin{array}{cc} 1 & -1 \\ 1 & 1 \end{array}\right) - 
\left(\begin{array}{cc} 1 & 0 \\ 0 & 1 \end{array}\right) \right],
\ee
the scattering amplitude in the basis $B$ thus takes the form (\ref{eq:MbasisDP}) with 
\be
{\cal M} = K \frac{e^{2i\delta_0}-1}{2i}.
\ee
This is precisely the expression given in \eq(\ref{eq:pw}).

\section{The fermion four-point function}

Using the plane-wave expansion of the quantum fields $\psi$ and $\bar\psi$, we can 
write the two-fermion correlation function as 
\ba
&&\Big\<\bar\psi_\gamma(t,\vec x')\, \psi_{\delta'}(t',\vec y')\; \psi_\alpha(0,-\vec p)\bar\psi_\beta(0,\vec p)\Big\>
\\ && \stackrel{t>t'>0}{=}
\frac{1}{2E_p} \sum_{\vec p'',\vec p'}\sum_{s,s',s'',s'''} e^{-E_{\vec p''}(t-t')}
\Big\<0\Big|a^{\bar F}_{\vec p''\,s''}   a_{\vec p's'''}^F\; e^{-Ht'} a^{\bar F}_{\vec p s} \;a^F_{(-\vec p) s'}\Big|0\Big\>
\nn && \qquad e^{i(\vec p''\cdot\vec x'+\vec p'\cdot\vec y')} \;
\frac{\bar v_\gamma^{s''}(\vec p'')\, u_{\delta '}^{s'''}(\vec p')
v_\alpha^s(\vec p) \bar u_\beta^{s'}(-\vec p) }{\sqrt{2E_{p''}2E_{p'}}}.
\nonumber
\ea
Now we insert a complete set of states in the remaining matrix element, the contribution of 
a bound state $|B\>$ is
\ba
 \<0|a^{\bar F}_{\vec p'' s''}  a^F_{\vec p' s'''}|B\>\, e^{-E_Bt'}
\,\<B | a^{\bar F}_{\vec p s} \;a^F_{(-\vec p) s'}|0\>.
\ea
The individual matrix elements are given by 
\ba
\<0|a_{\vec p''s''}^{\bar F} a_{\vec p's'''}^F |B\> &=& 
\frac{1}{\sqrt{2L^3}} \tilde\Psi(\vec p') \delta_{\vec p''(-\vec p')}
\left(\delta_{s''(-)} \delta_{s'''(+)} + \delta_{s''(+)} \delta_{s'''(-)} \right),
\qquad 
\\
\<B|  a^{\bar F}_{\vec p s} \;a^F_{(-\vec p) s'}|0\>
&=& \frac{1}{\sqrt{2L^3}}\; (-\tilde\Psi^*(-\vec p)) 
\left(\delta_{s'(+)} \delta_{s(-)} +\delta_{s'(-)} \delta_{s(+)} \right).
\ea
Finally, we obtain the expression for $t'\to\infty$
\ba
&&\Big\<\bar\psi_\gamma(t,\vec x')\, \psi_{\delta'}(t',\vec y')\; \psi_\alpha(0,-\vec p)\bar\psi_\beta(0,\vec p) \Big\> 
\\ && = \frac{-1}{L^3}\; \tilde\Psi(-\vec p)^* \; \frac{e^{- E_B t'}}{2\cdot2E_p}  
\left(v_\alpha^{(-)}(\vec p) \bar u_\beta^{(+)}(-\vec p) + v_\alpha^{(+)}(\vec p) \bar u_\beta^{(-)}(-\vec p)\right)
\nn && \qquad \sum_{\vec p''} \frac{e^{-E_{p''}(t-t')}}{2E_{p''}}\, e^{i\vec p''\cdot(\vec x'-\vec y')}
 \tilde \Psi(-\vec p'') \,
\left(\bar v_\gamma^{(-)}(\vec p'') u_{\delta'}^{(+)}(-\vec p'') + \bar v_\gamma^{(+)}(\vec p'') u_{\delta '}^{(-)}(-\vec p'')\right).
\nonumber
\ea
Equivalently, we can transform all fermion fields to the time-momentum representation and then obtain % \eq(\ref{eq:f4pt}).
\ba \la{eq:f4pt}
&&\Big\<\bar\psi_\gamma(t,\vec k)\, \psi_{\delta'}(t',-\vec q')\; \psi_\alpha(0,-\vec p)\bar\psi_\beta(0,\vec p) \Big\> 
\\ && = -L^3\;\delta_{\vec k\vec q'}\;  \tilde\Psi(-\vec p)^* \; \frac{e^{- E_B t'}}{2\cdot2E_p}  
\left(v_\alpha^{(-)}(\vec p) \bar u_\beta^{(+)}(-\vec p) + v_\alpha^{(+)}(\vec p) \bar u_\beta^{(-)}(-\vec p)\right)\cdot
\nn && \qquad  \frac{e^{-E_{\vec k}(t-t')}}{2E_{\vec k}}\,
 \tilde \Psi(-\vec k) \,
\left(\bar v_\gamma^{(-)}(\vec k) u_{\delta'}^{(+)}(-\vec k) + \bar v_\gamma^{(+)}(\vec k) u_{\delta '}^{(-)}(-\vec k)\right),
\nonumber
\ea

{\small 
\bibliographystyle{JHEP}
\bibliography{/Users/harvey/BIBLIO/viscobib.bib}
}
\end{document}